
%
%
\input harvmac

\def\Titlehh#1#2{\nopagenumbers\abstractfont\hsize=\hstitle\rightline{#1}%
\vskip .2in\centerline{\titlefont #2}\abstractfont\vskip .2in\pageno=0}
\def\CTPa{\it Center for Theoretical Physics, Department of Physics,
      Texas A\&M University}
\def\CTPb{\it College Station, TX 77843-4242, USA}
\def\HARCa{\it Astroparticle Physics Group,
Houston Advanced Research Center (HARC)}
\def\HARCb{\it The Woodlands, TX 77381, USA}

\def\CERN{\it CERN Theory Division, 1211 Geneva 23, Switzerland}
\def\ie{\hbox{\it i.e.}}     
\def\eg{\hbox{\it e.g.}}

\def\coeff#1#2{{\textstyle{#1\over #2}}}

\catcode`\@=11 

\def\lsim{\mathrel{\mathpalette\@versim<}}
\def\gsim{\mathrel{\mathpalette\@versim>}}
\def\@versim#1#2{\vcenter{\offinterlineskip
    \ialign{$\m@th#1\hfil##\hfil$\crcr#2\crcr\sim\crcr } }}
\def\boxit#1{\vbox{\hrule\hbox{\vrule\kern3pt
      \vbox{\kern3pt#1\kern3pt}\kern3pt\vrule}\hrule}}

\def\etal{{\it et. al.}}
\def\r#1{$\bf#1$}

\def\t1{{\tilde 1}}

\def\JL{J. L. Lopez}
\def\DVN{D. V. Nanopoulos}

\def\GeV{\,{\rm GeV}}
\def\TeV{\,{\rm TeV}}
\def\y{\,{\rm y}}

\def\wt{\widetilde}

\def\gluino{m_{\tilde g}}
\def\NPB#1#2#3{Nucl. Phys. B {\bf#1} (19#2) #3}
\def\PLB#1#2#3{Phys. Lett. B {\bf#1} (19#2) #3}

\def\PRD#1#2#3{Phys. Rev. D {\bf#1} (19#2) #3}
\def\PRL#1#2#3{Phys. Rev. Lett. {\bf#1} (19#2) #3}
\def\PRT#1#2#3{Phys. Rep. {\bf#1} (19#2) #3}

\def\TAMU#1{Texas A \& M University preprint CTP-TAMU-#1}

\nref\CAN{A. Chamseddine, R. Arnowitt, and P. Nath, \PRL{49}{82}{970};
For reviews see: R. Arnowitt and P. Nath, {\it Applied N=1 Supergravity}
(World Scientific, Singapore 1983); H. P. Nilles, \PRT{110}{84}{1}.}
\nref\EKN{J. Ellis, S. Kelley and D. V.  Nanopoulos, \PLB{249}{90}{441},
\PLB{260}{91}{131}; P. Langacker and M.-X. Luo, \PRD{44}{91}{817};
U. Amaldi, W. de Boer, and H. F\"urstenau, \PLB{260}{91}{447};
F. Anselmo, L. Cifarelli, A. Peterman, and A. Zichichi, Nuovo Cim. {\bf104A}
(1991) 1817.}
\nref\const{J. Ellis, S. Kelley, and \DVN, \NPB{373}{92}{55} and
\PLB{287}{92}{95}; R. Barbieri and L. Hall, \PRL{68}{92}{752}; F. Anselmo,
L. Cifarelli, A. Peterman, and A. Zichichi, Nuovo Cim. {\bf105A} (1992) 581;
J. Hisano, H. Murayama, and T. Yanagida, \PRL{69}{92}{1014}.}
\nref\WSY{S. Weinberg, \PRD{26}{82}{287}; N. Sakai and T. Yanagida,
\NPB{197}{82}{533}.}
\nref\ENR{J. Ellis, \DVN, and S. Rudaz, \NPB{202}{82}{43};
B. Campbell, J. Ellis, and \DVN, \PLB{141}{84}{229}.}
\nref\EMN{K. Enqvist, A. Masiero, and \DVN, \PLB{156}{85}{209}.}
\nref\CANpd{P. Nath, A. Chamseddine, and R. Arnowitt, \PRD{32}{85}{2348};
P. Nath and R. Arnowitt, \PRD{38}{88}{1479}.}
\nref\Japspd{M. Matsumoto, J. Arafune, H. Tanaka, and K. Shiraishi,
University of Tokyo preprint ICRR-267-92-5 (April 1992).}
\nref\ANabc{R. Arnowitt and P. Nath, \PRL{69}{92}{725}; P. Nath and
R. Arnowitt, \PLB{287}{92}{89} and NUB-TH-3048-92 and CTP-TAMU-27-92.}
\nref\MHTb{J. Hisano, H. Murayama, and T. Yanagida, Tohoku University preprint
TU--400 (July 1992).}
\nref\LNZ{\JL, \DVN, and A. Zichichi, \TAMU{49/92} and CERN-TH.6554/92.}
\nref\GRZ{G. Gamberini, G. Ridolfi, and F. Zwirner, \NPB{331}{90}{331}.}
\nref\EZ{J. Ellis and F. Zwirner, \NPB{338}{90}{317}.}
\nref\noscale{S. Kelley, \JL, \DVN, H. Pois, and K. Yuan, \PLB{273}{91}{423}.}
\nref\Japs{K. Inoue, M. Kawasaki, M. Yamaguchi, and T. Yanagida,
\PRD{45}{92}{328}; G. G. Ross and R. G. Roberts, \NPB{377}{92}{571};
M. Drees and M. M. Nojiri, \PRD{45}{92}{2482}.}
\nref\aspects{S. Kelley, \JL, \DVN, H. Pois, and K. Yuan, \TAMU{16/92} and
CERN-TH.6498/92.}
\nref\PDG{Particle Data Group, \PRD{45}{92}{S1}.}
\nref\KT{See \eg, E. Kolb and M. Turner, {\it The Early Universe}
(Addison-Wesley, 1990).}
\nref\inflation{For recent reviews see \eg, K. Olive, \PRT{190}{90}{307};
D. Goldwirth and T. Piran, \PRT{214}{92}{223}.}
\nref\LNY{\JL, \DVN, and K. Yuan, \NPB{370}{92}{445}.}
\nref\KM{M. Kawasaki and S. Mizuta, \PRD{46}{92}{1634};
S. Kelley, \JL, \DVN, H. Pois, and K. Yuan, \TAMU{56/92} and CERN-TH.6584/92;
M. Drees and M. M. Nojiri, SLAC preprint SLAC-PUB-5860 (July 1992).}
\nref\KZ{Z. Kunszt and F. Zwirner, CERN-TH.6150/91, ETH-TH/91-7.}
\nref\others{V. Barger, K. Cheung, A. L. Stange, and R. Phillips, MAD-PH-704,
June 1992; H. Baer, C. Kao, M. Bisset, X. Tata, and D. Dicus, FSU-HEP-920724.
For a recent review see J. Gunion, UCD-92-20.}
\nref\review{M. Chen, C. Dionisi, M. Martinez, and X. Tata,
\PRT{159}{88}{203}.}
\nref\Smoot{G. Smoot, \etal, COBE preprint (1992).}

\nfig\I{Scatter plot of the proton lifetime
$\tau_p\equiv\tau(p\to\bar\nu_{\mu,\tau}K^+)$ versus the gluino mass for the
hypercube of the parameter space explored. The current experimental lower bound
is $\tau^{exp}_p=1\times10^{32}\y$. The various `branches' correspond to fixed
values of $\xi_0$ as indicated (the labelling applies to all four windows).
The bottom row includes the cosmological constraint. The upper bound on
$\gluino$ follows from the requirement $m_{\tilde q}<1\TeV$.}
\nfig\II{The calculated neutralino relic density $\Omega_\chi h^2_0$ versus
the neutralino mass. Note the effect of the $s$-channel $Z$-pole at
$m_\chi\approx{1\over2}M_Z$. Only about $1/6$ of the points have
$\Omega_\chi h^2_0\le1$.}
\nfig\III{The calculated value of $\mu$ versus $\gluino$. The cosmological
constraint is enforced in the bottom row. Above the solid lines (labelled
$\wt B$) the neutralino is a nearly pure bino eigenstate, above the line
$|\mu|=M_2=0.3\gluino$ it is mostly a gaugino eigenstate. The small regions
in the bottom row plots which extend up into the pure bino regions correspond
to $m_\chi$ masses near ${1\over2}m_Z$ and ${1\over2}m_h$.}
\nfig\IV{The lightest neutralino mass ($m_\chi$) versus the gluino mass
(top row) and the second-to-lightest neutralino mass ($m_{\chi^0_2}$) versus
the lightest chargino mass ($m_{\chi^+_1}$) (bottom row). These plots show the
anticipated approximate mass correlations due to the proton decay constraint,
$m_\chi\approx0.15\gluino$, $m_{\chi^0_2}\approx m_{\chi^+_1}\approx2m_\chi$.
The cosmological constraint (not enforced in this figure) simply depletes the
point density, without affecting the range of the particle masses.}
\nfig\V{The one-loop corrected lightest Higgs boson mass ($m_h$) versus the
lightest chargino mass. If the cosmological constraint is enforced (bottom
row), then for $m_h\gsim80\GeV$ the lightest chargino is below $110\GeV$.
Thus, at least one of these particles is likely to be discovered at LEPII.
A significant number of the allowed points correspond to $m_\chi$ masses near
${1\over2}M_Z$ and ${1\over2}m_h$.}

\Titlehh{\vbox{\baselineskip12pt\hbox{CERN-TH.6628/92}
\hbox{CTP--TAMU--61/92}\hbox{ACT--19/92}}}
{\vbox{\centerline{Proton Decay and Cosmology Strongly Constrain}
\vskip2pt\centerline{the Minimal SU(5) Supergravity Model}}}
\centerline{JORGE~L.~LOPEZ$^{(a)(b)}$\footnote{$^\dagger$}{A condensed version
of this paper will appear in the Proceedings of the XXVI International
Conference on High Energy Physics, Dallas--Texas, August 5--12, 1992.
Conference Speaker.}, D.~V.~NANOPOULOS$^{(a)(b)(c)}$, and H. POIS$^{(a)(b)}$}
\smallskip
\centerline{$^{(a)}$\CTPa}
\centerline{\CTPb}
\centerline{$^{(b)}$\HARCa}
\centerline{\HARCb}
\centerline{$^{(c)}$\CERN}
\vskip .1in
\centerline{ABSTRACT}
We present the results of an extensive exploration of the five-dimensional
parameter space of the minimal $SU(5)$ supergravity model, including the
constraints of a long enough proton lifetime ($\tau_p>1\times10^{32}\y$) and
a small enough neutralino cosmological relic density ($\Omega_\chi h^2_0\le1$).
We find that the combined effect of these two constraints is quite severe,
although still leaving a small region of parameter space with
$m_{\tilde g,\tilde q}<1\TeV$. The allowed values of the proton lifetime
extend up to $\tau_p\approx1\times10^{33}\y$ and should be fully explored by
the SuperKamiokande experiment. The proton lifetime cut also entails the
following mass correlations and bounds: $m_h\lsim100\GeV$,
$m_\chi\approx{1\over2}m_{\chi^0_2}\approx0.15\gluino$,
$m_{\chi^0_2}\approx m_{\chi^+_1}$, and $m_\chi<85\,(115)\GeV$,
$m_{\chi^0_2,\chi^+_1}<165\,(225)\GeV$ for $\alpha_3=0.113\,(0.120)$.
Finally, the {\it combined} proton decay and cosmology constraints predict
that if $m_h\gsim75\,(80)\GeV$ then $m_{\chi^+_1}\lsim90\,(110)\GeV$ for
$\alpha_3=0.113\,(0.120)$. Thus, if this model is correct, at least one of
these
particles will likely be observed at LEPII.
\bigskip
{\vbox{\baselineskip12pt\hbox{CERN-TH.6628/92}\hbox{CTP--TAMU--61/92}
\hbox{ACT--19/92}}}
\Date{August, 1992}

\newsec{Introduction}
The minimal $SU(5)$ supergravity model \CAN\ has recently passed the simplest
possible consistency check, namely the unification of the gauge couplings
at an energy scale $M_U\sim10^{16}\GeV$ \EKN. This check depends only in
sub-leading order on the masses of the light and heavy particles in the theory,
and as such provides weak constraints on the various model parameters \const.
On the other hand, the requirement that the dimension-five--induced proton
decay operators \WSY\ be within current experimental bounds provide rather
stringent constraints on all sectors of the theory
\refs{\ENR,\EMN,\CANpd,\Japspd,\ANabc,\MHTb}. Recently two of us (JL and DVN)
with A. Zichichi \LNZ, studied a representative set of points in parameter
space which satify the proton decay bound and applied to these the cosmological
requirement $\Omega_\chi h^2_0\le1$, where $\Omega_\chi$ is the relic abundance
of the lightest neutralino (also the lightest supersymmetric particle (LSP) and
which is assumed to be stable) and $0.5\le h_0\le1$ is the Hubble parameter.
We found that the cosmological constraint was grossly violated for these
points.
It was also noted that there may still exist cosmologically allowed regions for
sufficiently small values of $m_t$. In this paper we present a systematic
exploration of the five-dimensional parameter space of the model, which
corroborates the previous indicative results. We identify the small regions
of parameter space which satisfy both constraints and show that the
cosmologically acceptable region should be almost fully testable at LEPII
and/or SuperKamiokande.

The minimal $SU(5)$ supergravity model consists of the Standard Model fields
(with two Higgs doublets) and their superpartners, plus the heavy GUT fields
in the form of colored Higgs triplet fields ($H,\bar H$) and gauge and Higgs
bosons in the \r{24} representation of $SU(5)$. Universal soft-supersymmetry
breaking at the scale $M_U$ is described by three parameters: $m_{1/2},m_0,A$.
The $SU(5)$ gauge symmetry entails gauge coupling ($\alpha_1=\alpha_2=\alpha_3
=\alpha_5$ at $M_U$) and Yukawa coupling ($\lambda_b=\lambda_\tau$ at $M_U$)
unification. In our calculation all these parameters are run from $M_U$ down to
low energies ($M_Z$) as prescribed by the relevant renormalization group
equations. Finally, at the scale $M_Z$ the full one-loop effective potential is
minimized and the one-loop corrected Higgs boson masses are obtained. The final
parameter list is: $\tan\beta$, $m_t$, $m_{1/2}$, $\xi_0\equiv m_0/m_{1/2}$,
$\xi_A\equiv A/m_{1/2}$, and the sign of the Higgs mixing term ($\mu$). This
set of parameters should be constrasted with the 21-dimensional parameter space
of the usual MSSM. For recent analyses of this type, see Refs.
\refs{\GRZ,\EZ,\noscale,\Japs,\aspects}.

\newsec{Standard Constraints}
We impose the following set of `standard constraints' on the parameter space
of the model (see Ref. \aspects\ for a detailed discussion):
\item{(i)} one-loop radiative electroweak symmetry breaking;
\item{(ii)} perturbative unification (which implies $m_t\lsim190\GeV$ and
$\tan\beta\lsim50$);
\item{(iii)}$m^2_{\tilde q,\tilde l}>0$;
\item{(iv)} a neutral and colorless LSP (\ie, $\tilde\nu$ or
$\chi\equiv\chi^0_1$);
\item{(v)} experimental bounds on $m_{\chi^+_1},m_{\tilde l},
m_{\tilde g},m_{\tilde q},m_t,\Gamma_{inv}$;
\item{(vi)} $m^{ave}_{\tilde q},m_{\tilde g}<1\TeV$, motivated by naturalness
considerations, or demanding testability of the model at the SSC;
\item{(vii)}$\lambda_b(M_U)=\lambda_\tau(M_U)$, which in practice we use to
predict $m_b(m_b)$ for a given set of $m_t$, $\tan\beta$, and $\alpha_3$
values.
\medskip
We have explored the following hypercube of the parameter space:
$\mu>0,\mu<0$, $\tan\beta=2-10\,(2)$, $m_t=100-160\,(5)$, $\xi_0=0-10\,(1)$,
$\xi_A=-\xi_0,0,+\xi_0$, and $m_{1/2}=50-300\,(6)$, where the numbers in
parenthesis represent the size of the step taken in that particular direction.
(Points outside these ranges have little (a posteriori) likelihood of being
acceptable.) Of these $92,235\times2=184,470$ points, $\approx25\%$ passed
all the standard constraints. In what follows we present our results as a
collection of scatter plots, where a given pair of observables is plotted for
each allowed point in the parameter space.

\newsec{Proton Decay}
In unified supersymmetric theories only the dimension-five--mediated proton
decay operators are constraining. In calculating the proton lifetime we
consider the typically dominant decay modes $p\to \bar\nu_{\mu,\tau} K^+$ and
neglect all other possible modes. Schematically the lifetime is given by \CANpd
\eqn\I{\tau_p\equiv\tau(p\to\bar\nu_{\mu,\tau}K^+)
\sim\left| M_H \sin2\beta {1\over f}{1\over 1+y^{tK}}\right|^2.}
Here $M_H$ is the mass of the exchanged GUT Higgs triplet which on perturbative
grounds is assumed to be bounded above by $M_H<3M_U$ \refs{\EMN,\ANabc,\MHTb};
$\sin2\beta=2\tan\beta/(1+\tan^2\beta)$,
thus $\tau_p$ `likes' small $\tan\beta$ (we find that only $\tan\beta\le6$ is
allowed); $1+y^{tK}$ represents the calculable ratio of the third- to the
second-generation contributions to the dressing one-loop diagrams. An unkown
phase appears in this ratio and we always consider the weakest possible case of
destructive interference. Finally $f$ represents the sparticle-mass--dependent
dressing one-loop function which decreases asymptotically with large sparticle
masses.

In Fig. 1 (top row) we show a scatter plot of $(\tau_p,m_{\tilde g})$. The
various `branches' correspond to fixed values of $\xi_0$. Note that for
$\xi_0<3$, $\tau_p<\tau^{exp}_p=1\times10^{32}\y$ (at 90\% C.L. \PDG). Also,
for a given value of
$\xi_0$, there is a corresponding allowed interval in $\gluino$. The lower end
of this interval is determined by the fact that $\tau_p\propto1/f^2$, and
$f\approx m_{\chi^+_1}/m^2_{\tilde q}\propto 1/\gluino(c+\xi^2_0)$, in the
proton-decay--favored limit of $\mu\gg M_W$; thus
$\gluino(c+\xi^2_0)>{\rm constant}$. The upper end of the interval follows from
our requirement $m_{\tilde q}(\propto \gluino\sqrt{c+\xi^2_0})<1\TeV$.
Statistically speaking, the proton decay cut is quite severe,
allowing only about $\sim1/10$ of the points which passed all the standard
constraints, independently of the sign of $\mu$.

Note that if we take $M_H=M_U$ (instead of $M_H=3M_U$), then
$\tau_p\to{1\over9}\tau_p$ and all points in Fig. 1 would become excluded.
To obtain a rigorous lower bound on $M_H$, we would need to explore the lowest
possible allowed values of $\tan\beta$ (in Fig. 1, $\tan\beta\ge2$). Roughly,
since the dominant $\tan\beta$ dependence of $\tau_p$ is through the explicit
$\sin2\beta$ factor, the upper bound $\tau_p\lsim8\times10^{32}\y$ for
$\tan\beta=2$, would become $\tau_p\lsim1\times10^{33}\y$ for $\tan\beta=1$.
Therefore, the current experimental lower bound on $\tau_p$ would imply
$M_H\gsim M_U$. Note also that SuperKamiokande
($\tau^{exp}_p\approx2\times10^{33}\y$) should be able to probe the whole
allowed range of $\tau_p$ values.

The actual value of $\alpha_3(M_Z)$ used in the calculations ($\alpha_3=0.120$
in all figures) has a non-negligible effect of some of the final results,
mostly due to its effect on the value of $M_H=3M_U$: larger values of
$\alpha_3$
increase $M_U$ and therefore $\tau_p$, and thus open up the parameter space,
and viceversa. For example, for $\alpha_3=0.113\,(0.120)$ we get
$\gluino\lsim550\,(800)\GeV$, $\xi_0\ge5\,(3)$, and
$\tau_p\lsim4\,(8)\times10^{32}\y$.

\newsec{Cosmology}
We assume that the lightest neutralino is a stable particle, as expected in
the minimal, $R$-parity conserving model. The current cosmological observations
of $\Omega_0\lsim1$ \KT\ and/or the inflation prediction $\Omega_0=1$
\inflation, lead us to impose the constraint $\Omega_\chi h^2_0\le1$. In Fig. 2
we show the calculated values\foot{For a detailed discussion of the methods
used to compute $\Omega_\chi h^2_0$ see Ref. \LNY. For computations of
$\Omega_\chi h^2_0$ in supergravity models with radiative electroweak symmetry
breaking see \refs{\EZ,\noscale,\KM}.} of $\Omega_\chi h^2_0$ versus $m_\chi$,
which show a noticeable dip at $m_\chi\approx{1\over2}M_Z$ due to $s$-channel
$Z$-pole annihilation. Only $\sim1/6$ of the points have
$\Omega_\chi h^2_0\le1$. This result is not unexpected since proton decay is
suppressed by heavy sparticle masses, whereas $\Omega_\chi h^2_0$ is enhanced.
Therefore, a delicate balance needs to be attained to satisfy both constraints
simultaneously. The subset of cosmologically
allowed points does not change the range of possible $\tau_p$ values (see
Fig. 1 bottom row), although it depletes the constant-$\xi_0$ `branches'.

In Fig. 3 we show the calculated values of $\mu$ with and without the
cosmological constraint. As expected, values of $\mu$ giving nearly pure
gaugino $\chi$-compositions (above the $|\mu|=M_2=0.3\gluino$ line) are
cosmologically disfavored since in this limit the $\chi\chi Z$ and $\chi\chi h$
couplings are highly suppressed. Exceptions to this rule can be traced to
$\gluino$ values for which $m_\chi\approx{1\over2}M_Z,{1\over2}m_h$ when the
annihilation cross section is enhanced by $s$-channel poles. Also, since
$|\mu|$
grows with $m_t$, larger values of $m_t$ tend to be disfavored as well. The
results are quite similar for the $\alpha_3=0.113$ case.

\newsec{Particle Mass Correlations}
Since Fig. 3 shows that proton decay generally requires $\mu\gg M_W$ (and to a
somewhat lesser extent also $\mu\gg M_2$), the lightest chargino will have mass
$m_{\chi^+_1}\approx M_2\approx0.3\gluino$, whereas the two lightest
neutralinos
will have masses $m_\chi\approx M_1\approx{1\over2}M_2$ and
$m_{\chi^0_2}\approx M_2$ \ANabc. Thus, within some approximation we
expect
\eqn\II{m_\chi\approx\coeff{1}{2}m_{\chi^0_2},\qquad
m_{\chi^0_2}\approx m_{\chi^+_1},\qquad m_\chi\approx0.15\gluino.}
The calculated values of these masses are shown in Fig. 4 (without imposing the
cosmological constraint). The approximate mass relations are quite accurate for
$\mu>0$, but more qualitative for $\mu<0$. Inclusion of the cosmological
constraint basically just depletes the point density without affecting
significantly the range of particle masses. The value of $\alpha_3$ does not
affect these mass relations either, although the particle mass ranges do change
\eqn\III{m_\chi<85\,(115)\GeV,\qquad m_{\chi^0_2,\chi^+_1}<165\,(225)\GeV,
\qquad{\rm for}\ \alpha_3=0.113\,(0.120).}
We also find that the one-loop corrected lightest Higgs boson mass ($m_h$) is
bounded above by
\eqn\IV{m_h\lsim110\GeV,}
independently of the sign of $\mu$, the value of $\alpha_3$, or the
cosmological constraint. In Fig. 5 we plot $m_h$ versus $m_{\chi^+_1}$
which shows an experimentally interesting correlation {\it if} the
cosmological constraints are imposed (bottom row). Indeed,
\eqn\V{m_h\gsim75\,(80)\GeV \Rightarrow m_{\chi^+_1}\lsim90\,(110)\GeV,}
for $\alpha_3=0.113\,(0.120)$. Moreover,
the correlations among the lightest chargino and neutralino masses in Eq. \II\
imply analogous results for $(m_h,m_{\chi^0_2})$ and $(m_h,m_\chi)$,
\eqn\VI{m_h\gsim80\GeV \Rightarrow m_{\chi^0_2}\lsim90\,(110)\GeV,
\quad m_\chi\lsim48\,(60)\GeV,}
for $\alpha_3=0.113\,(0.120)$.
These correlations can be understood in the following way: since we find that
$m_A\gg M_Z$, then $m_h\approx|\cos2\beta|M_Z+({\rm rad.\ corr.})$. In the
situation we consider here, we have determined that all of the allowed points
for $m_{\tilde g}>400\;\GeV$ correspond to $\tan\beta=2$. This implies that
the tree-level contribution to $m_h$ is $\approx55\GeV$.\foot{Note that Fig. 5
cannot be used to establish a lower bound on $m_h$ since values of
$1<\tan\beta<2$ have not been considered. This is unlike the case for the
quoted upper bound in Eq. \IV.} We also find that the cosmology cut restricts
$m_t<130(140)\;\GeV$ for $\mu>0(\mu<0)$ in this range of $\gluino$. Therefore,
the radiative correction contribution to $m^2_h$ ($\propto m^4_t$) will be
modest in this range of $\gluino$. This explains the depletion of points for
$m_h\gsim 80\;\GeV$ in Fig. 5 and leads to the mass relationships in Eqs.
\V\ and \VI.

Recent studies of Higgs searches relevant to LEPII have shown that large
regions of the parameter space which determine $m_h$, including
radiative corrections, can be explored \refs{\KZ,\others}.
These studies make simplifying assumptions regarding the many parameters at
low energy, as well as choosing fixed values of $m_t$. Nonetheless, we expect
a certain level of quantitative agreement with these generic analyses.
Important conclusions tend to be unanimous; if $\tan\beta\lsim 5$, then
values of $m_h$ up to $m_h\approx 80 \GeV$ can be explored (see \eg,
Figs. 2a,8a in \KZ). This is precisely the constraint on $\tan\beta$ that is
realized in our analysis due to the stringent proton decay cuts.
As we have discussed, the regions in parameter space where $m_h\gsim 80 \GeV$
result in the constraint $m_{\chi^+_1}\lsim90\,(110)\GeV$ for
$\alpha_3=0.113\,(0.120)$, and both signs of $\mu$.
Early model-dependent analyses of $e^+e^-\rightarrow {\wt W}l{\tilde \nu_l}$
indicate that $m_{\wt W}\lsim 100\GeV$ could be explored at LEPII \review.
A more careful upper limit would require a detailed calculation, but these
results are encouraging nonetheless. Therefore, if LEPII does not see the
lightest Higgs, it has a good chance of seeing the lightest chargino instead,
or viceversa.

\newsec{Conclusions}
The most direct and pervasive evidence for unified models would be the
observation of nucleon decay. In fact, this kind of test of a unified model
has the very appealing property of involving the physics of both low-mass
and high-mass particles in the theory, and as such should be able to
discriminate among the various competing unified models at hand. In the
specific
case of the minimal $SU(5)$ supergravity model, we have shown that under
sensible assumptions, the range of proton lifetimes still to be probed by
the next round of proton decay experiments is finite and completely accessible.
Besides the various uncertainties on the measured `constants' which enter the
proton lifetime formula, two assumptions are key to the results presented in
this paper: (i) the upper bound on the Higgs triplet mass $M_H<3M_U$, and (ii)
the assumed upper bound on the squark and gluino masses of $1\TeV$. Relaxing
any of these assumptions can suppress the proton lifetime to acceptable values.
However, we believe that if these quite sensible assumptions had to be relaxed
to ensure compatibility with experimental requirements, then much of the
motivation to consider this model would fade away and in effect the model
would become `sociologically' excluded. Another important variable in this
study is the value of $\alpha_3$, larger values of which tend to open up
the parameter space. We have considered the case of $\alpha_3=0.113$ and 0.120.
A better defined value for this quantity will go a long way in nailing
the specific predictions of the model.

The recent precise observation by the COBE satellite of minute anisotropies
in the cosmic microwave background radiation \Smoot\ is a dramatic reminder
that precision cosmology has come to age, and that a unified model should be
judged seriously by its consistency with cosmological constraints. In the
present model we have shown that a suitable value for the cosmological relic
density of the lightest neutralino is another powerful constraint on the model
parameters. In effect it reduces the size of the allowed parameter space by a
factor of 6.

The proton decay cut singles out a small region of the parameter space where a
number of fairly accurate mass correlations and mass bounds exists. We find for
the one-loop corrected lightest Higgs boson mass $m_h\lsim110\GeV$,
independently
of the sign of $\mu$, the value of $\alpha_3$, or the cosmological constraint.
Since $|\mu|\gg M_W$ holds for most of the allowed parameter space, it also
follows that $m_\chi\approx{1\over2}m_{\chi^0_2}\approx0.15\gluino$ and
$m_{\chi^0_2}\approx m_{\chi^+_1}$. The imposed upper bound $m_{\tilde
q}<1\TeV$
cuts off the value of the gluino mass for a given $\xi_0$ value and gives
upper bounds on the light particle masses: $m_\chi<85\,(115)\GeV$ and
$m_{\chi^0_2,\chi^+_1}<165\,(225)\GeV$ for $\alpha_3=0.113\,(0.120)$.
The effect of the cosmological cut on these predictions is negligible.
However, this cut does remove a large portion of the allowed points and
has a dramatic effect on the correlation between $m_h$ and $m_{\chi^+_1}$
(or $m_{\chi^0_2}, m_\chi$ since they are all related), such that at least
one of these particles is quite likely to be observable at LEPII.
To recapitulate, of the 184,470 points in parameter space which we examined,
only $\sim1/240$ satisfy the standard constraints, the proton decay bound, and
the cosmological requirement, leaving only a rather restricted set of points
to be put to experimental test. In sum, it should not be long before we could
start discriminating among the various supersymmetric unified models.

\bigskip
\bigskip
\noindent{\it Acknowledgments}: This work has been supported in part by DOE
grant DE-FG05-91-ER-40633. The work of J.L. has been supported in part by an
ICSC-World Laboratory Scholarship and in part by an SSC Fellowship. The work of
D.V.N. has been supported in part by a grant from Conoco Inc. We would like to
thank the HARC Supercomputer Center for the use of their NEC SX-3
supercomputer.
\listrefs
\listfigs
\bye